\documentclass{PoS}

\usepackage{epsfig}
\usepackage{amsmath}
\usepackage{latexsym}
\usepackage{amssymb}
\usepackage{dsfont}

\newcommand{\ud}{\mathrm{d}}

\newcommand{\uTr}{\mathrm{Tr}}

\title{Generalized Transverse-Momentum Dependent  Parton Distributions in Light-Cone Quark Models}

\ShortTitle{GTMDs in Light-Cone Quark Models}

\author{\speaker{Cédric Lorcé}\\
				Institut für Kernphysik, Johannes Gutenberg-Universität,\\ 
				Mainz, D-55099, Germany\\
        E-mail: \email{lorce@kph.uni-mainz.de}}

\author{Barbara Pasquini\\
				Universit\`a degli Studi di Pavia,
        Dipartimento di Fisica Nucleare e Teorica \\
				and INFN, Sezione di Pavia,
				Pavia, I-27100, Italy\\
        E-mail: \email{Barbara.Pasquini@pv.infn.it}}

\abstract{We discuss the general formalism for the calculation in light-cone quark models of the  fully unintegrated, off-diagonal  quark-quark 
correlator of the nucleon.
The corresponding distributions in impact parameter space are the Wigner or phase-space distributions. The results obtained in two different light-cone quark models in the case of unpolarized quarks in an unpolarized proton are very similar and present a non-trivial shape which can be understood as due to the orbital motion of the quarks.}

\FullConference{Light Cone 2010 - LC2010\\
		June 14-18, 2010\\
		Valencia, Spain}

\begin{document}

\section{Introduction}

Hadrons are composite objects. Their interaction with external probes like \emph{e.g.} photons is parameterized in terms of Lorentz-invariant functions. The most complete information is contained in the so-called generalized transverse-momentum dependent parton distributions (GTMDs). After appropriate Fourier 
transform, the GTMDs can be interpreted as Wigner or phase-space distributions, giving access to the correlations between quark momentum and transverse position.

A brief introduction to the generalized quark-quark correlator functions defining the GTMDs can be found in Section~\ref{Sec2}. On the light cone, this correlator can be written as the overlap of light-cone wave functions. We present in Section~\ref{Sec3} the overlap restricted to the three quark (3Q) sector and specify the expression to a light-front constituent quark model (LFCQM) and the chiral quark-soliton model ($\chi$QSM). In Section~\ref{Sec4} it is explained why working on the light cone is mandatory to develop a (quasi-)probabilistic interpretation of the distributions. Finally, we apply in Section~\ref{Sec5} the formalism presented in the previous sections to study the distribution of an unpolarized quark in an unpolarized proton.

\section{General Quark-Quark Correlator}\label{Sec2}

\begin{figure}[ht]
\begin{center}
\epsfig{file=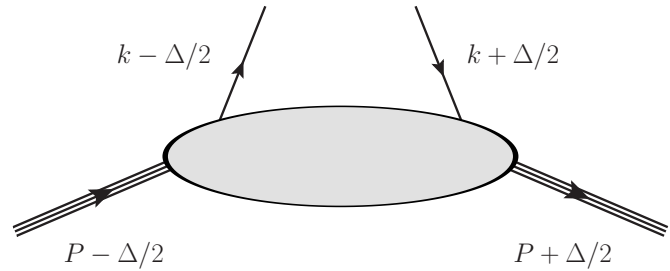,  width=0.6\columnwidth}
\end{center}
\caption{Kinematics for the fully-unintegrated quark-quark correlator.}
\label{fig1}
\end{figure}

The maximum amount of information on the partonic structure of the nucleon is contained in the fully-unintegrated quark-quark correlator $\tilde W$ for a spin-$1/2$ hadron~\cite{Ji:2003ak,Belitsky:2003nz,Belitsky:2005qn,Meissner:2009ww}, defined as
\begin{equation}\label{gencorr}
\tilde W^{[\Gamma]}_{\Lambda'\Lambda}(P,k,\Delta,N;\eta)=\frac{1}{2}\int\frac{\ud^4z}{(2\pi)^4}\,e^{ik\cdot z}\,\langle p',\Lambda'|\overline\psi(-\tfrac{1}{2}z)\Gamma\,\mathcal W\,\psi(\tfrac{1}{2}z)|p,\Lambda\rangle.
\end{equation}
This correlator is a function of the initial and final hadron light-cone helicities $\Lambda$ and $\Lambda'$, the average hadron and quark four-momenta $P=(p'+p)/2$ and $k$, and the four-momentum transfer to the hadron $\Delta=p'-p$ (see Fig.~\ref{fig1} for the kinematics).
 The superscript $\Gamma$ stands for any element of the basis $\{\mathds 1,\gamma_5,\gamma^\mu,\gamma^\mu\gamma_5,i\sigma^{\mu\nu}\}$ in Dirac space. A Wilson line $\mathcal W\equiv\mathcal W(-\tfrac{1}{2}z,\tfrac{1}{2}z|n)$ ensures the color gauge invariance of the correlator, connecting the points $-\tfrac{1}{2}z$ and $\tfrac{1}{2}z$ \emph{via} the intermediary points $-\tfrac{1}{2}z+\infty\cdot n$ and $\tfrac{1}{2}z+\infty\cdot n$ by straight lines. This induces a dependence of the Wilson line on the light-cone direction $n$. Since any rescaled four-vector $\alpha n$ with some positive parameter $\alpha$ could be used to specify the Wilson line, the correlator actually only depends on the four-vector $N=\frac{M^2n}{P\cdot n}$, where $M$ is the hadron mass. The parameter $\eta=\textrm{sign}(n^0)$ gives the sign of the zeroth component of $n$, \emph{i.e.} indicates whether the Wilson line is future-pointing ($\eta=+1$) or past-pointing ($\eta=-1$). 

The quark-quark correlators parametrized in terms of generalized parton distributions (GPDs), tranverse-momentum dependent parton distributions (TMDs) and form factors (FFs) correspond to specific limits or projections of Eq.~\eqref{gencorr}. These correlators have in common the fact that the quark fields are taken at the same light-cone time $z^+=0$. Let us then focus our attention on the $k^-$-integrated version of Eq.~\eqref{gencorr}
\begin{equation}\label{GTMDcorr}
\begin{split}
W^{[\Gamma]}_{\Lambda'\Lambda}(P,x,\vec k_\perp,\Delta,N;\eta)&=\int\ud k^-\,\tilde W^{[\Gamma]}_{\Lambda'\Lambda}(P,k,\Delta,N;\eta)\\
&=\frac{1}{2}\int\frac{\ud z^-\,\ud^2z_\perp}{(2\pi)^3}\,e^{ik\cdot z}\,\langle p',\Lambda'|\overline\psi(-\tfrac{1}{2}z)\Gamma\,\mathcal W\,\psi(\tfrac{1}{2}z)|p,\Lambda\rangle\Big|_{z^+=0},
\end{split}
\end{equation}
where we used for a generic four-vector $a^\mu=[a^+,a^-,\vec a_\perp]$ the light-cone components $a^\pm=(a^0\pm a^3)/\sqrt{2}$ and the transverse components $\vec a_\perp=(a^1,a^2)$, and $x=k^+/P^+$ is the average fraction of longitudinal momentum carried by the quark. A complete parametrization of this object in terms of so-called generalized transverse momentum dependent parton distributions (GTMDs) has been achieved in~\cite{Meissner:2009ww}. The GTMDs can be considered as the \emph{mother distributions} of GPDs and TMDs. For example, the (T-even) distribution of unpolarized quarks in an unpolarized hadron is given by the GTMD $F^e_{11}$, and is related to the GPD $H$ and the TMD $f_1$ as follows
\begin{equation}
\label{eq:f1}
\begin{split}
H(x,0,\vec\Delta_\perp^2)&=\int\textrm{d}^2k_\perp\,F^e_{11}(x,0,\vec k_\perp^2,\vec k_\perp\cdot\vec\Delta_\perp,\vec\Delta_\perp^2),\\
f_1(x,\vec k_\perp^2)&=F^e_{11}(x,0,\vec k_\perp^2,0,0).
\end{split}
\end{equation}

\section{Overlap Representation}\label{Sec3}

Following the lines of~\cite{Diehl:2000xz,Brodsky:2000xy}, we obtain in the light-cone gauge $A^+=0$ an overlap representation for the correlator~\eqref{GTMDcorr} at the twist-two level restricted to the 3Q Fock sector\footnote{Quark flavor and color indices have been omitted for clarity. In the processes considered here the flavor and color of a given quark remain unchanged.} 
\begin{equation}\label{overlap}
W^{[\Gamma]}_{\Lambda'\Lambda}(P,x,\vec k_\perp,\Delta,N;\eta)=\frac{1}{\sqrt{1-\xi^2}}\sum_{\lambda'_i,\lambda_i}\int[\ud x]_3\,[\ud^2k_\perp]_3\,\Delta(\tilde k)\,\psi^*_{\Lambda'\beta'}(r')\,\psi_{\Lambda\beta}(r)\prod_{i=1}^3 M^{\lambda'_i\lambda_i},
\end{equation}
where the integration measures are defined as
\begin{equation}
[\ud x]_3\equiv\left[\prod_{i=1}^3\ud x_i\right]\delta\!\!\left(1-\sum_{i=1}^3x_i\right),\qquad[\ud^2k_\perp]_3\equiv\left[\prod_{i=1}^3\frac{\ud^2k_{i\perp}}{2(2\pi)^3}\right]2(2\pi)^3\,\delta^{(2)}\!\!\left(\sum_{i=1}^3\vec k_{i\perp}\right).
\end{equation}
Furthermore, in Eq.~\eqref{overlap} the function $\Delta(\tilde k)=3\,\Theta(x_1)\,\delta(x-x_1)\,\delta^{(2)}(\vec k_\perp-\vec k_{1\perp})$ selects the active quark average momentum (we choose to label the active quark with $i=1$ and the spectator quarks with $i=j=2,3$). The 3Q LCWF $\psi_{\Lambda\beta}(r)$ depends on the momentum coordinates $\tilde k_i=(y_i,\kappa_{i\perp})$ of the quarks relative to the hadron momentum (collectively indicated by $r$), and the index $\beta$ which stands for the set of the quark light-cone helicities $\{\lambda_i\}$. The transition from the initial quark light-cone helicity $\lambda_i$ to the final one $\lambda'_i$ is described by a complex-valued $2\times 2$ matrix $M^{\lambda'_i\lambda_i}$. In particular, we have for the spectator quarks $M^{\lambda'_j\lambda_j}=\delta^{\lambda'_j\lambda_j}$. For the active quark, the matrix $M^{\lambda'_1\lambda_1}$ depends on the twist-two Dirac structure $\Gamma_\text{twist-2}=\{\gamma^+,\gamma^+\gamma_5,i\sigma^{+1}\gamma_5,i\sigma^{+2}\gamma_5\}$ used in the correlator, see \emph{e.g.}~\cite{Boffi:2002yy,Boffi:2003yj,Pasquini:2005dk}.

We choose to work in an infinite momentum frame such that $P^+$ is large, $\vec P_\perp=\vec 0_\perp$ and $\Delta\cdot P=0$. The four-momenta involved are then
\begin{equation}
\begin{aligned}
P&=\left[P^+,\frac{M^2+\tfrac{\Delta_\perp^2}{4}}{2(1-\xi^2)P^+},\vec 0_\perp\right],\qquad
&\Delta&=\left[-2\xi P^+,\xi\,
\frac{M^2+\tfrac{\Delta_\perp^2}{4}}{(1-\xi^2)P^+},\vec\Delta_\perp\right],\\
k&=\left[xP^+,k^-,\vec k_\perp\right],&n&=\left[0,\pm 1,\vec 0_\perp\right].
\end{aligned}
\end{equation}
Note that the form used for $n$ is not the most general one, but leads to an appropriate definition of TMDs for SIDIS and DY processes. For the active and spectator quarks the initial and final momentum coordinates are then
\begin{equation}
\begin{aligned}
\tilde k_1&=\left(\frac{x+\xi}{1+\xi},\vec k_\perp-\frac{1-x}{1+\xi}\,\frac{\vec\Delta_\perp}{2}\right),\qquad&\tilde k'_1&=\left(\frac{x-\xi}{1-\xi},\vec k_\perp+\frac{1-x}{1-\xi}\,\frac{\vec\Delta_\perp}{2}\right),\\
\tilde k_j&=\left(\frac{x_j}{1+\xi},\vec k_{j\perp}+\frac{x_j}{1+\xi}\,\frac{\vec\Delta_\perp}{2}\right),&\tilde k'_j&=\left(\frac{x_j}{1-\xi},\vec k_{j\perp}-\frac{x_j}{1-\xi}\,\frac{\vec\Delta_\perp}{2}\right).
\end{aligned}
\end{equation}

So far, the exact 3Q LCWF derived directly from the QCD Lagrangian is not known. Nevertheless, we can try to reproduce the gross features of hadron structure at low  scales using constituent quark models. Many models exist on the market based on the concept of constituent quarks. However only a few incorporate consistently relativistic effects. We focus here on two such models: the light-front constituent quark model (LFCQM)~\cite{Boffi:2002yy,Boffi:2003yj,Pasquini:2005dk} and the chiral quark-soliton model ($\chi$QSM)~\cite{Petrov:2002jr,Diakonov:2005ib,Lorce:2006nq,Lorce:2007as,Lorce:2007fa}. The LCWFs used in LFCQM and in $\chi$QSM have a very similar structure given by
\begin{equation}\label{LCWF}
\psi_{\Lambda\beta}(r)=\Psi(r)\sum_{\sigma_i}\Phi_\Lambda^{\sigma_1\sigma_2\sigma_3}\prod_{i=1}^3 D_{\lambda_i\sigma_i}(\tilde k_i),
\end{equation}
where $\Psi(r)$ is a global symmetric momentum wave function, $\Phi_\Lambda^{\sigma_1\sigma_2\sigma_3}$ is the $SU(6)$ spin-flavor wave function, and $D(\tilde k)$ is an $SU(2)$ matrix connecting light-cone helicity $\lambda_i$ and canonical spin $\sigma_i$
\begin{equation}\label{generalizedMelosh}
D(\tilde k)=\frac{1}{|\vec K|}\begin{pmatrix}K_z&K_L\\-K_R&K_z\end{pmatrix},\qquad K_{R,L}=K^1\pm iK^2.
\end{equation}
The explicit expressions for the momentum wave function $\Psi(r)$ 
in Eq.~\eqref{LCWF} and the vector $\vec K$ in Eq.~\eqref{generalizedMelosh} 
in LFCQM read
\begin{equation}
\Psi(r)=2(2\pi)^3\sqrt{\frac{\omega_1\omega_2\omega_3}{x_1x_2x_3\mathcal M_0}}\,\frac{\mathcal N}{(\mathcal M_0^2+\beta^2)^\gamma},\qquad K_z=m+y\mathcal M_0,\qquad \vec K_\perp=\vec\kappa_\perp,\qquad \kappa_z=y\mathcal M_0-\omega,
\end{equation}
where $\mathcal N$ is a normalization factor, $\mathcal M_0=\sum_i\omega_i$ is the free invariant mass, $\omega_i$ is the free energy of quark $i$, $m$ is the constituent quark mass, and $\beta,\gamma$ are  model parameters fitted to reproduce the anomalous magnetic moments of the nucleon~\cite{Pasquini:2007iz}. On the other hand, within the $\chi$QSM one has
\begin{equation}
\Psi(r)=\mathcal N\prod_{i=1}^3|\vec K_i|,\qquad K_z=h+\frac{\kappa_z}{|\vec\kappa|}\,j,\qquad \vec K_\perp=\frac{\vec\kappa_\perp}{|\vec\kappa|}\,j,\qquad \kappa_z=y\mathcal M_N-E_\text{lev},
\end{equation}
where $\mathcal M_N$ is the soliton mass, $E_\text{lev}$ is the energy of the discrete level in the spectrum, and $h,j$ are the upper and lower components of the Dirac spinor describing this discrete level.

For further convenience we introduce the tensor correlator
\begin{equation}
W^{\mu\nu}\equiv\frac{1}{2}\uTr\left[\bar\sigma^\mu W^\nu\right]=\frac{1}{2}\sum_{\Lambda'\Lambda}(\bar\sigma^\mu)^{\Lambda\Lambda'}W^\nu_{\Lambda'\Lambda},
\end{equation}
where $W^\nu_{\Lambda'\Lambda}\equiv\left(W^{[\gamma^+]}_{\Lambda'\Lambda},W^{[i\sigma^{+1}\gamma_5]}_{\Lambda'\Lambda},W^{[i\sigma^{+2}\gamma_5]}_{\Lambda'\Lambda},W^{[\gamma^+\gamma_5]}_{\Lambda'\Lambda}\right)$ and $\bar\sigma^\mu=(\mathds{1},\vec \sigma)$ with $\sigma_i$ the Pauli matrices. We now use the LCWF given by Eq.~\eqref{LCWF} and write the overlap representation of the correlator tensor $W^{\mu\nu}$ as
\begin{equation}\label{master1}
W^{\mu\nu}(P,x,\vec k_\perp,\Delta,N;\eta)=\frac{1}{\sqrt{1-\xi^2}}\int[\ud x]_3\,[\ud^2k_\perp]_3\,\Delta(\tilde k)\,\Psi^*(r')\,\Psi(r)\,\mathcal A^{\mu\nu}(r',r),
\end{equation}
where $\mathcal A^{\mu\nu}(r',r)$ stands for 
\begin{equation}\label{master2}
\mathcal A^{\mu\nu}(r',r)=A\,O_1^{\mu\nu}\left(l_2\cdot l_3\right)+B\left[l_2^\mu\left(l_3\cdot O_1\right)^\nu+l_3^\mu\left(l_2\cdot O_1\right)^\nu\right].
\end{equation}
In Eq.~\eqref{master2}, $l^\mu_j=O^{\mu0}_j$ and the matrix $O^{\mu\nu}$ is given by
\begin{equation}\label{spinhelicity}
O^{\mu\nu}=\frac{1}{|\vec K'||\vec K|}\begin{pmatrix}
\vec K'\cdot\vec K&i\left(\vec K'\times\vec K\right)_x&i\left(\vec K'\times\vec K\right)_y&-i\left(\vec K'\times\vec K\right)_z\\
i\left(\vec K'\times\vec K\right)_x&\vec K'\cdot\vec K-2K'_xK_x&-K'_xK_y-K'_yK_x&K'_xK_z+K'_zK_x\\
i\left(\vec K'\times\vec K\right)_y&-K'_yK_x-K'_xK_y&\vec K'\cdot\vec K-2K'_yK_y&K'_yK_z+K'_zK_y\\
i\left(\vec K'\times\vec K\right)_z&-K'_zK_x-K'_xK_z&-K'_zK_y-K'_yK_z&-\vec K'\cdot\vec K+2K'_zK_z
\end{pmatrix}.
\end{equation}
The tensor correlator $W^{\mu\nu}$ in Eq. (3.9) has two indices. The index 
$\mu$ refers to the transition in terms of baryon light-cone helicity, while the index $\nu$ refers to the transition in terms of the active quark light-cone helicity. For example, the components $W^{00}$ and $W^{03}$ correspond to the matrix elements of the $\gamma^+$ and $\gamma^+\gamma_5$ operators in the case of an unpolarized hadron, respectively. Equation~\eqref{master1} gives the explicit expression for the tensor correlator in terms of the overlap of initial $\Psi(r)$ and final $\Psi^*(r')$ symmetric (instant-form) momentum wave functions with the tensor $\mathcal A^{\mu\nu}(r',r)$ for a fixed mean momentum of the active quark $\tilde k$. 
The tensor $\mathcal A^{\mu\nu}(r',r)$ contains the spin-flavor structure 
derived from the overlap of the three initial and final quarks.
Taking into account the possible couplings of the helicities of the active and spectator quarks to give the hadron helicity, the coefficient 
$A$ and $B$ in Eq.~\eqref{master2}  for SU(6)  spin-flavor wave functions
are \begin{equation}
A^p_u=4,\qquad B^p_u=1,\qquad A^p_d=-1,\qquad B^p_d=2.
\end{equation}
Furthermore, the matrix $O^{\mu\nu}$ in Eq.~(3.12) describes the overlap 
of the initial and final quark states. The columns are labeled by the index $\nu$ which indicates the type of transition in terms of quark light-cone helicity. The rows are labeled by the index $\mu$ which indicates the type of transition in terms of quark canonical spin.
This matrix reduces to
$l^\mu_i=O^{\mu0}_i$ for the spectator quarks, 
since in this case the light-cone helicity is conserved.

\section{Impact Parameter Space}\label{Sec4}

According to the standard interpretation \cite{Ernst:1960zza,Sachs:1962zzc}, the charge density can be identified with the three-dimensional Fourier transform of the electric Sachs Form Factor $G_E$, \emph{i.e.} the matrix element of the current density $J^0$
\begin{equation}
\rho(\vec r)=\int\frac{\textrm{d}^3q}{(2\pi)^3}\,e^{-i\vec q\cdot\vec r}\,G_E(Q^2).
\end{equation}
This identification is actually only valid in the nonrelativistic approximation. To work out the Fourier transform, one has to know the form factors for every $Q^2$. In the Breit frame the latter is identified with the three-momentum of the virtual photon $Q^2=\vec q\,^2$. This means that for every value of $Q^2$ we have to move to a different frame and the charge density undergoes naturally a different Lorentz contraction. Moreover, in order to have a probabilistic/charge density interpretation, the number of particles should be conserved. 
However, in the Breit frame nothing prevents the photon to create or annihilate a quark-antiquark pair.

All these problems are cured in the infinite momentum frame with $q^+=0$ (the so-called Drell-Yan-West frame). The photon is kinematically not allowed to change the number of quarks as the light-cone momentum of a massive particle is strictly positive $p^+>0$. Moreover, the hadron undergoes an extreme Lorentz contraction and looks like a pancake. Only a two-dimensional charge density \cite{Soper:1976jc,Burkardt:2000za,Burkardt:2002hr,Miller:2007uy,Carlson:2007xd} is then meaningful and can be identified with the two-dimensional Fourier transform of the matrix element of $J^+$
\begin{equation}
\rho_{\vec s}(\vec b_\perp)=\int\frac{\textrm{d}^2q_\perp}{(2\pi)^2}\,\frac{e^{-i\vec q_\perp\cdot\vec b_\perp}}{2p^+}\,\langle p^+,\tfrac{\vec q_\perp}{2},\vec s|J^+(0)|p^+,-\tfrac{\vec q_\perp}{2},\vec s\rangle,
\end{equation}
the photon virtuality being given by $Q^2=\vec q_\perp\,\!\!\!\!^2$, and $\vec s$ denoting the hadron polarization.

\section{Wigner Distributions}\label{Sec5}

Wigner distributions are quantum phase-space distributions, containing all the correlations between position and momentum of the partons. Since Heisenberg's uncertainty principle forbids to determine precisely both position and momentum of a quantum state, Wigner distributions have to be considered as quasi-probabilistic distributions. In the context of quantum field theory, they have already been discussed to some extent in the Breit frame \cite{Belitsky:2003nz}. For the reason mentioned in the previous section, it is actually preferable to work in the infinite momentum frame. By Fourier transforming $F^e_{11}$ in Eq.\eqref{eq:f1} with respect to $\vec\Delta_\perp$ and integrating over $x$, we obtain a (transverse) phase-space distribution $\rho(\vec k_\perp^2,\vec k_\perp\cdot\vec b_\perp,\vec b^2_\perp)$. The only two available transverse vectors are $\vec k_\perp$ and $\vec b_\perp$. This means that for fixed $\hat k_\perp\cdot\hat b_\perp$ and $|\vec k_\perp|$ the distribution is axially symmetric, which is physically meaningful since there is not a preferred direction  in the transverse plane and 
a global rotation around $\hat e_z$ should not have any effect on the  distribution
 of unpolarized quarks in an unpolarized nucleon.

If we are now interested in the amplitude to find a fixed transverse momentum $\vec k_\perp$ in the transverse plane, the distribution is not axially symmetric anymore due to the $\vec k_\perp\cdot\vec b_\perp$ dependence. Although the Wigner distributions can not be directly extracted from experiments, they can be inferred from the LCWFs parametrized on 
the basis of our phenomenological knowledge of GPDs, TMDs and form factors.
As a starting point, we used the  $\chi$QSM and LFCQM which were introduced 
in Section~\ref{Sec3} and already tested in the calculation of 
 several nucleon observables \cite{Lorce:2006nq,Lorce:2007as,Lorce:2007fa,Pasquini:2007iz,Boffi:2009sh}.

\begin{figure}
\begin{center}
\begin{tabular}{c@{\hspace{1cm}}c}
\psfig{file=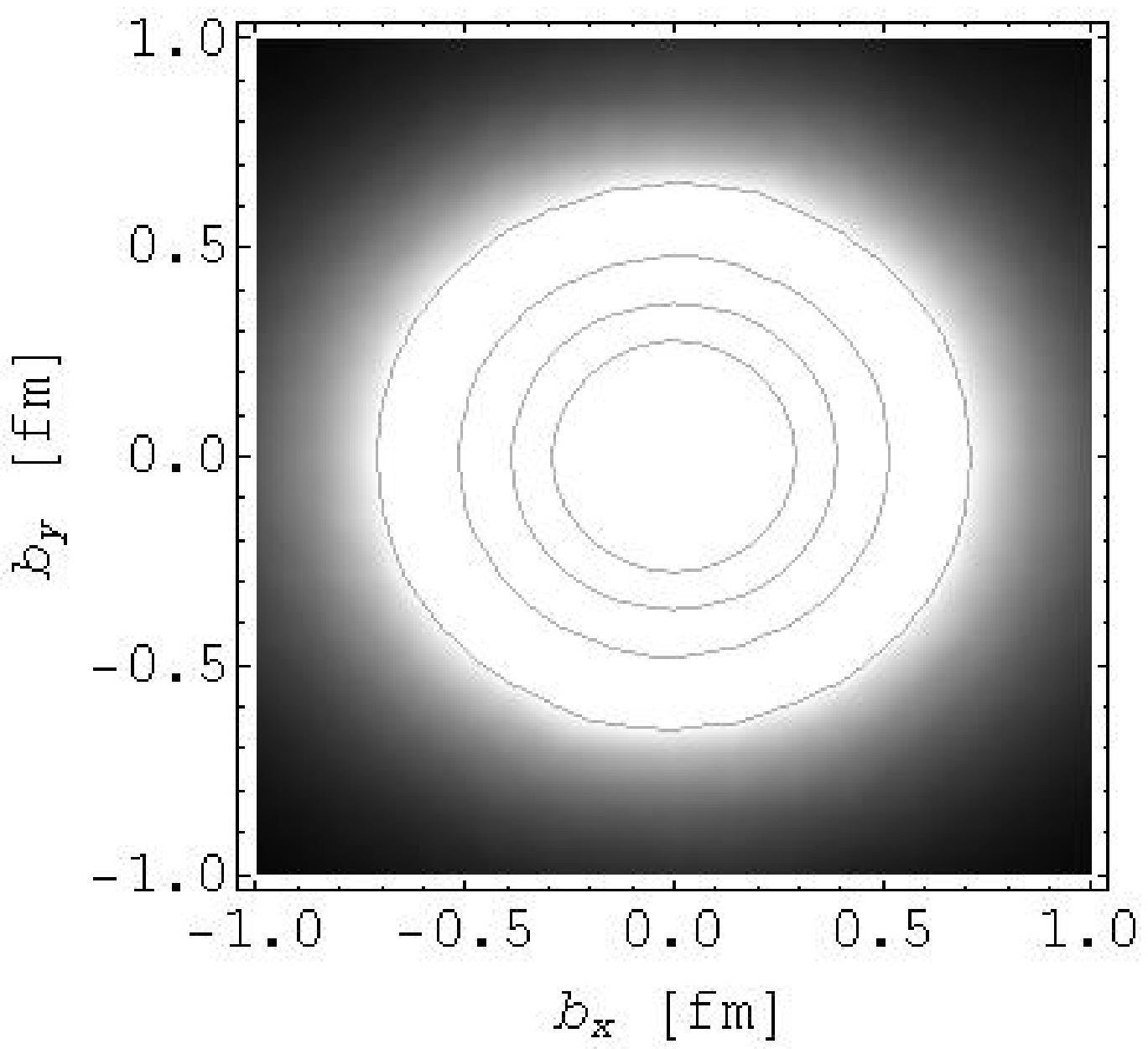,width=5cm}&\psfig{file=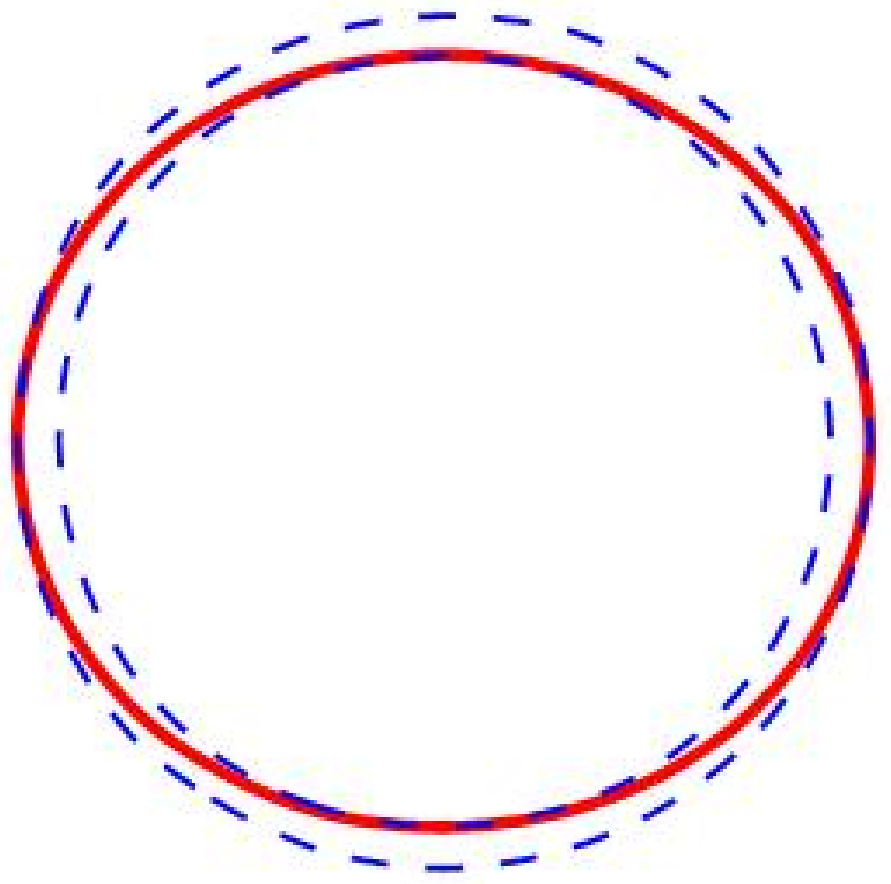,width=4cm}
\end{tabular}
\end{center}
\caption{Typical distribution (left) and equi-amplitude line (right) of fixed $\vec k_\perp=k_\perp\,\hat e_y$ in the transverse plane for an unpolarized quark in an unpolarized nucleon.}
\label{aba:fig1}
\end{figure}

In the case of $F^e_{11}$, both models give the same qualitative picture with a larger distribution amplitude when $k_\perp \perp b_\perp$ and smaller when $k_\perp \parallel b_\perp$ (see Fig.~\ref{aba:fig1}).
This can be understood with naive semi-classical arguments. The radial momentum $(\vec k_\perp\cdot\hat b_\perp)\,\hat b_\perp$ of a quark has to decrease rapidly in the periphery because of confinement. The polar momentum $\vec k_\perp-(\vec k_\perp\cdot\hat b_\perp)\,\hat b_\perp$ receives a contribution from the orbital motion of the quark which can still be significant in the periphery (in an orbital motion, one does not need to reduce the momentum to avoid a quark escape). This naive picture also tells us that this phenomenon should become more pronounced as we go to peripheral regions ($|\vec b_\perp|\gg$) and to high quark momenta ($|\vec k_\perp|\gg$). This tendency is supported by both models.

\begin{figure}
\begin{center}
\psfig{file=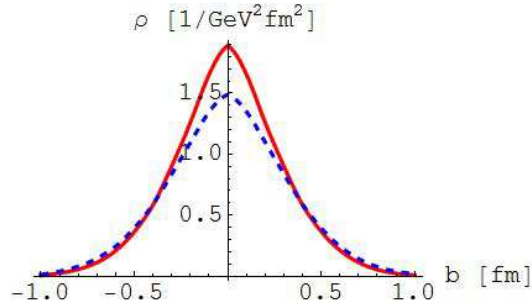,width=7cm}
\end{center}
\caption{Typical radial distributions in the transverse plane for fixed $\hat k_\perp\cdot\hat b$ and $|\vec k_\perp|$ of an unpolarized quark in an unpolarized proton. Solid line corresponds to half of the up quark distribution. Dashed line corresponds to the down quark distribution.}
\label{aba:fig2}
\end{figure}

It is also interesting to compare up and down quark distributions. Let us fix once again $\hat k_\perp\cdot\hat b_\perp$ and $|\vec k_\perp|$. The distribution being axially symmetric in the transverse plane, we focus on the radial distribution, see Fig.~\ref{aba:fig2}.
The up quark distribution has been divided by two for comparison with the down quark distribution. Up quarks appear to be more concentrated around the center than down quarks. For a neutron, we just have to exchange up and down quarks. We can therefore see that the center of the neutron is negative, in agreement with the conclusion obtained using the phenomenological neutron FFs \cite{Miller:2007uy}.

\newpage
\section{Summary}

We used the framework of light-cone wave function to study generalized transverse-momentum dependent parton distributions which parametrize the most general quark-quark correlator. These distributions are connected to the Wigner or phase-space distributions by Fourier transform. We presented a general expression for the 3Q contribution to the generalized correlation functions defining the GTMDs and applied it to a light-front constituent quark model and the chiral quark-soliton model. A non-trivial pattern for the phase-space distribution in the transverse plane for an unpolarized quark in an unpolarized proton has been observed and interpreted semi-classically as related to the quark orbital angular momentum. We also confirmed the picture of a negatively charged core in the neutron. A presentation of the systematic study of generalized transverse-momentum dependent parton distributions is in preparation.

\end{document}